\documentclass[preprint,14pt]{elsarticle}

\makeatletter
\def\ps@pprintTitle{%
    \let\@oddhead\@empty
    \let\@evenhead\@empty
    \def\@oddfoot{\reset@font\hfil\thepage\hfil}%
    \let\@evenfoot\@oddfoot}
\makeatother

\date{}

\usepackage{graphicx}
\usepackage{caption}
\usepackage{subcaption}
\usepackage{color}
\usepackage[bookmarks=false]{hyperref}
\usepackage{multirow}
\usepackage{amsmath,amssymb,amsfonts}
\usepackage{amsthm}
\usepackage{mathrsfs}
\usepackage{xcolor}
\usepackage{textcomp}
\usepackage{manyfoot}
\usepackage{booktabs}
\usepackage{algorithm}
\usepackage{algorithmicx}
\usepackage{algpseudocode}
\usepackage{listings}
\usepackage{siunitx}
\usepackage{float}

\usepackage{lineno}

\newcounter{bla}

\begin{document}

\begin{frontmatter}



 \title{KinetiX: A performance portable code generator for chemical kinetics and transport properties}




\author[a]{Bogdan A. Danciu\corref{cor1}}
\author[a]{Christos E. Frouzakis}

\affiliation[a]{organization={CAPS Laboratory, Department of Mechanical and Process Engineering, ETH Zürich},
               city={Zürich},
               postcode={8092},
               country={Switzerland}}

\cortext[cor1]{Corresponding author}

\begin{abstract}
We present \texttt{KinetiX}, a software toolkit to generate computationally efficient fuel-specific routines for the chemical source term, thermodynamic and mixture-averaged transport properties for use in combustion simulation codes. The C++ routines are designed for high-performance execution on both CPU and GPU architectures. 
On CPUs, chemical kinetics computations are optimized by eliminating redundant operations and using data alignment and loops with trivial access patterns that enable auto-vectorization, reducing the latency of complex mathematical operations. 
On GPUs, performance is improved by loop unrolling, reducing the number of costly exponential evaluations and limiting the number of live variables for better register usage. The accuracy of the generated routines is checked against reference values computed using Cantera and the maximum relative errors are below $10^{-7}$. We evaluate the performance of the kernels on some of the latest CPU and GPU architectures from AMD and NVIDIA, i.e., AMD EPYC 9653, AMD MI250X, and NVIDIA H100. The routines generated by \texttt{KinetiX} outperform the general-purpose Cantera library, achieving speedups of up to 2.4x for species production rates and 3.2x for mixture-averaged transport properties on CPUs. Compared to the routines generated by PelePhysics (CEPTR), \texttt{KinetiX} achieves speedups of up to 2.6x on CPUs and 1.7x on GPUs for the species production rates kernel on a single-threaded basis.


{\bf PROGRAM SUMMARY}

\begin{small}
\noindent
{\em Program Title:} KinetiX \\
{\em Developer's repository link:} \url{https://github.com/bogdandanciu/KinetiX} \\
{\em Licensing provisions:} BSD 2-clause  \\
{\em Programming language: } Python, C++ \\
{\em Nature of problem:} Combustion simulations require efficient computation of chemical source terms, thermodynamic and transport properties for diverse fuel types. The challenge is optimizing these computations for both CPUs and GPUs without compromising accuracy. \\
{\em Solution method:} Starting from an input file containing kinetic parameters, thermodynamic and transport data, \texttt{KinetiX} generates fuel-specific routines to compute species production rates, thermodynamic and mixture-averaged transport properties for high-performance execution on both CPU and GPU architectures.\\
\end{small}

\end{abstract}


\end{frontmatter}


\section{Introduction} \label{sec:introduction}


High fidelity Direct Numerical Simulations (DNS) have emerged as an invaluable tool for the study of combustion processes \cite{Poinsot}. DNS resolves both the turbulent flow and the complex reaction chemistry down to the smallest time and length scales, requiring enormous computational resources. The rapid development of High Performance Computing (HPC) has enabled DNS for novel fuels to investigate flames in turbulent flows. Although combustion simulation tools are becoming increasingly powerful, DNS of reactive flows has been mostly limited to relatively small domains and canonical geometries compared to the scale and complexity of real systems that would be beneficial for developing the next generation of carbon-neutral technologies. For applications involving  realistic engine configurations or practical combustor geometries, DNS has mainly been used for non-reactive simulations \cite{Fang2023, Danciu2023, Danciu2024}. Nevertheless, DNS results provide invaluable insights into the complex combustion phenomena and complement experimental data that is often limited by the extreme conditions present in advanced combustion applications, such as high pressures and temperatures.

Combustion chemistry is described by complex reaction mechanisms, which, depending on the fuel molecule size, contain tens to thousands of chemical species participating in roughly 5x as many reactions \cite{LU2009192}.
As a result, a considerable fraction of computational time in DNS can be invested in the evaluation of the species production rates, thermodynamic and transport properties \cite{Chen2011}. For decades, the general-purpose, problem-independent, transportable, FORTRAN chemical kinetics code package CHEMKIN \cite{Kee2017} developed in the 1970's at Sandia National Laboratory defined the standard for specifying chemical kinetic reaction mechanisms, thermochemical data, and transport properties, and provided a set of flexible and powerful tools for incorporating complex chemical kinetics into reacting flow simulations. The source code was available to the combustion community at no cost \cite{NAP13049}. 
When CHEMKIN became a for-profit product in 1997, Dave Goodwin started developing from scratch the open source code Cantera \cite{Cantera}, which uses the object oriented programming paradigm and offers new physical models and multiple programming interfaces (Matlab, Python, C/C++ and Fortran 90). Both codes aim at automating the incorporation of detailed chemical reaction mechanisms, thermodynamic properties and transport coefficients into combustion codes. 

Recent trends in HPC for scientific applications have seen a shift towards heterogeneous architectures, leveraging the massively-parallel processing capabilities of GPUs alongside traditional CPUs to accelerate computationally intensive tasks. Optimizing the computation of species production rates, thermodynamic and transport properties on heterogeneous computing platforms adds another layer of complexity. Computing the reaction rates involves expensive exponential operations for each chemical reaction. On CPUs, these long latency instructions can significantly reduce throughput if they are not properly vectorized. The massively-parallel processing capabilities of GPUs can be used to better hide these long-latency operations, but obtaining significant parallelism hinges on the fact that more threads need to remain active. This implies that the register pressure should remain low so that global memory access is avoided. On the other hand, the computational cost of evaluating transport properties can scale quadratically with the number of species requiring hundreds of live variables per grid point. Modern CPUs have a relatively large cache that can accommodate these large working sets. On GPUs, however, they often exceed the small on-chip memory allocated to each thread, resulting in register spilling, low occupancy, and underutilization of mathematical units.

Motivated by the potential reduction in computational costs, several tools have been developed to generate optimized kernels for different computing platforms. Zirwes et al. \cite{Zirwes2018, Zirwes2019} introduced a converter that translates an input file containing the reaction mechanisms into C++ source code. The generated code allows the reaction mechanisms to be restructured for efficient computation while generating densely packed data with linear access patterns that can be vectorized to exploit maximum performance on CPU systems. The tool is, however, limited to computing the species production rates and it is only optimized for CPUs. Bauer et al.~\cite{Bauer2014} developed Singe, a Domain Specific Language (DSL) compiler for combustion chemistry that leverages warp specialization to produce high performance code for NVIDIA GPUs. 
PelePhysics \cite{PeleSoftware} provides CEPTR, a tool that supports the integration of chemical models specified in Cantera YAML format, offering greater flexibility and cross-platform optimization. CEPTR parses the reaction mechanism file and outputs an optimized C++ library with routines to compute species production rates, and thermodynamic properties, as well as the physical parameters required to compute species- and temperature-dependent molecular transport coefficients. 

Due to the chemical stiffness exhibited by combustion kinetics, many solvers typically rely on robust, high-order implicit integration algorithms based on backward differentiation (BDF) formulas \cite{Curtiss1952, Byrne1987, Brown1989, Hindmarsh2005}. The chemical stiffness can be compounded by diffusive, convective, and acoustic phenomena, and an operator-split formulation is commonly used to reduce the integration of chemical source terms to a zero-dimensional setting \cite{Najm}.
In order to solve the nonlinear algebraic equations that arise in BDF methods, the Jacobian matrix or its product with a vector must be evaluated \cite{Hindmarsh2005}. Several software tools have been developed that in addition to the routines for the source terms offer generation of their analytical Jacobian. pyJac \cite{Niemeyer2017} is a Python-based open-source program that generates analytical Jacobian matrices for use in chemical kinetics modeling and analysis. In addition, pyJac uses an optimized evaluation order to minimize computational and memory operations, which is optimized for CPUs and GPUs through the NVIDIA CUDA framework \cite{Nickolls2008}. 
TChem \cite{Kim2023} is a portable software toolkit for the analysis of complex combustion mechanisms. The software offers tools for gas-phase and surface chemistry, thermodynamic properties as well as analytic Jacobians computed through automatic differentiation. TChem uses the Kokkos framework \cite{Trott2022} to achieve portability across multiple heterogeneous computing platforms with a single version of the code.

Our use cases for the optimized kernels are two highly-efficient spectral element solvers for the DNS of low Mach number combustion. The first one is the plugin LAVp \cite{SKphd,brambilla,behrooz} based on the CFD solver Nek5000 \cite{nek5000} and targets CPU HPC systems, while its successor nekCRF \cite{nekCRF} is developed for the GPU-accelerated NekRS \cite{nekRS}. In LAVp, thermochemistry and mixture-averaged transport was mainly handled by general-purpose CHEMKIN routines. Fuel-specific optimized thermochemistry routines generated by Fuego \cite{Fuego} have also been used. For the new reactive flow plugin, nekCRF, the need for a more efficient library capable of handling heterogeneous computing platforms became apparent. 

The closest open-source software toolkit that could meet the requirements of nekCRF, providing routines for both CPUs and GPUs, is the PelePhysics source code generator CEPTR. However, it does not generate routines for mixture-averaged transport properties, which are calculated within PelePhysics. In addition, CEPTR generates the pure species coefficients using third-order polynomial fits following the CHEMKIN approach, while Cantera uses fourth-order polynomial fits. 
\texttt{KinetiX} was developed to address these limitations and meet the specific requirements of nekCRF.
The starting point is a Cantera YAML file that contains the reaction mechanism with its kinetic parameters as well as thermodynamic and transport data. The input is parsed by a converter to generate C++ source code files containing all the necessary functions to compute the thermochemical and transport properties, designed for efficient execution on CPUs and GPUs. Although \texttt{KinetiX} was originally developed for use in nekCRF, the routines generated are general enough to be coupled with other combustion codes.

The rest of the paper is organized as follows. Section \ref{sec:code_generation} describes the techniques used to optimize the evaluation of the routines on CPUs and GPUs. Next, in Sec.~\ref{sec:results}, we discuss the tools available in \texttt{KinetiX}, and Secs.~\ref{subsec:validation} and~\ref{subsec:performance} demonstrate the correctness and computational performance of the routines generated for benchmark chemical kinetic models on some of the latest CPU and GPU architectures from AMD and NVIDIA i.e., AMD EPYC 9654, AMD MI250X and NVIDIA H100. We also discuss the implications of these results in these sections. The main conclusions and future research directions are outlined in Sec.~\ref{sec:conclusions}.
For completeness, \ref{sec:theory} summarizes the expressions employed for the reaction rates, thermodynamic and mixture-averaged transport properties. 

\section{Optimized code generation} \label{sec:code_generation}

\subsection{Basic concept of the code generation approach} \label{subsec:generator_concept}


The code generator produces fuel-specific efficient routines to compute the chemical production rates, thermodynamics and mixture-averaged transport properties on CPUs and GPUs. The Cantera YAML file that provides the thermodynamic and transport data together with the detailed reaction mechanism is parsed by the converter to generate C++ source files. Since the computed quantities only depend on the local mixture properties, the optimization needs to be done at the node level. For GPU parallelization, \texttt{KinetiX} employs grid-level parallelism, assigning one thread per grid point to compute all quantities.

\begin{figure}[H]
    \centering
    \includegraphics[width=\linewidth]{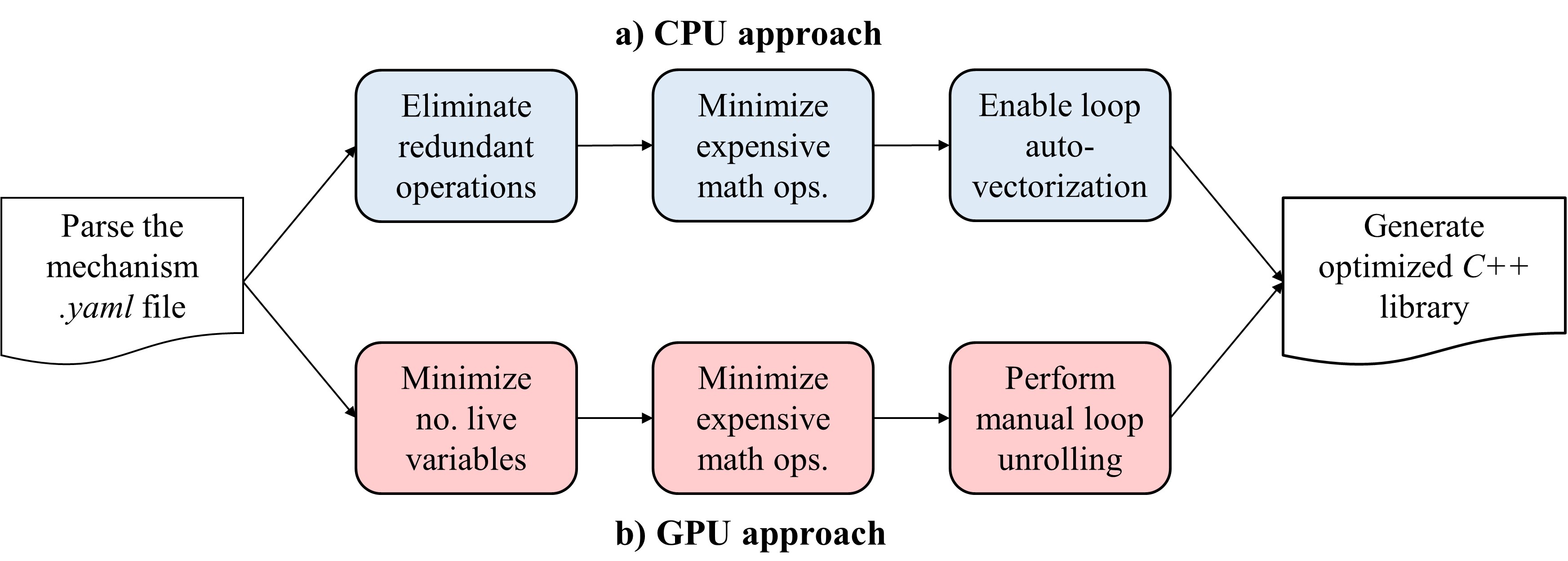}
    \caption{Simplified overview over how the code generator works including the two main optimization approaches.}
    \label{fig:workflow}
\end{figure}

Two main optimization approaches have been adopted to generate routines that are better suited for execution on CPUs or GPUs shown schematically in Fig.~\ref{fig:workflow} and can be summarized as follows:
\begin{itemize}
  \item
    The first approach is particularly efficient on CPUs and is motivated by the work of Zirwes et al.~\cite{Zirwes2018, Zirwes2019}. The generated C++ routines optimize the computation of chemical rates by first restructuring and reordering the reaction mechanism so that redundant operations resulting from already computed reaction rate constants are avoided, and reactions are grouped according to their type to optimize the reuse of cached results and minimize code branching. In addition, data is aligned in memory and loops with trivial access patterns are generated to facilitate auto-vectorization of grouped reactions, which better hides the long latency mathematical instructions. For mixture-averaged transport properties, the coefficients for the polynomial fits of the pure species properties are stored densely packed in memory and are further defined as compile-time constants, which together with loop optimization strategies enable auto-vectorization to compute the mixture-averaged transport properties more efficiently. 
  \item
    The second approach is particularly efficient on GPUs. Since these have relatively small memory caches and registers, it is performance critical that the register pressure of the kernels, which occurs when not enough registers are available for a given task, remains low. By avoiding slow global memory accesses, more threads remain active concurrently, increasing occupancy and effectively hiding the increased latency of computationally intensive mathematical operations like $\exp$ and $\log$. One of the main goals in the code generator is therefore to minimize the number of live variables within GPU execution. The previous approach of grouping rate constants and reactions together, while efficient on CPUs, poses a potential challenge, since it requires storing many intermediate values, which can significantly increase register pressure. Instead, the progress rates of each reaction is computed separately while also minimizing redundant operations and expensive mathematical instructions without increasing register pressure. For the mixture-averaged transport properties, the loops to compute the pure species properties are manually unrolled to avoid large working sets for the coefficients of the polynomial fits, which scale quadratically with the number of species in the computation of species diffusivities. 
\end{itemize}

In the following sections, the two code generation approaches used in \texttt{KinetiX} are explained in more detail. The focus is on examples for forward and reverse rate constants and mixture-averaged diffusivities. Similar optimization strategies are also applied to other generated properties, which are not discussed further here.

\subsection{Code generation strategy for CPUs}

\subsubsection{Forward and reverse rate constants} \label{subsec:cpu_rates}

The first step in computing the species production rates is to determine the forward rate constants expressed using Arrhenius law as  
$k_f = A\,T^{\beta}\,\exp(-E_a/\mathcal{R}T)$, where $A$ is the pre-exponential factor, $\beta$ the temperature exponent, $E_a$ the activation energy, and $\mathcal{R}$ the ideal gas constant (Eq.~(\ref{eq:arrhenius})). For a detailed reaction mechanism such as Konnov's scheme \cite{Konnov2009} for ethanol (C$_2$H$_5$OH) with 129 species and 1231 reactions, the naive approach requires evaluating the exponential 1231 times. As described in ~\ref{sec:fallof}, each of the 37 falloff reaction requires the computation of two rate constants $k_0$ and $k_{\infty}$ to obtain the $k_f$, bringing the total number of exponential evaluations based on Arrhenius law to 1268.

In the Cantera implementation, all forward rate constants are computed by calling the exponential function irrespective of the values of the Arrhenius parameters $A$, $\beta$ and $E_a$. Two of the special cases for $\beta=E_a=0$ and $\beta$ an integer and $E_a=0$ are highlighted in Eq.~\ref{eq:k_f}. Another case would be the reuse of already computed rate constants when $\beta$ and $E_a$ are the same for two reactions. For the C$_2$H$_5$OH mechanism, the three distinct cases can be summarized as follows:
\begin{itemize}
    \item For 347 rate constants, $\beta=E_a=0$ and $k_f=A$. 
    \item For another 7 rate constants, $E_a=0$ and $\beta$ is a small non-zero integer, $k_f=A T^{\beta}$ and $T^{\beta}$ can be computed using multiplications instead of exponentiation. 
    \item If different rate constants share the same values of $\beta$ and $E_a$, the exponential is computed only once and reused. This occurs 264 times.  
\end{itemize}

\begin{figure}[ht]
    \begin{center}
    \makebox[\textwidth]{%
        \includegraphics[width=1.15\linewidth]{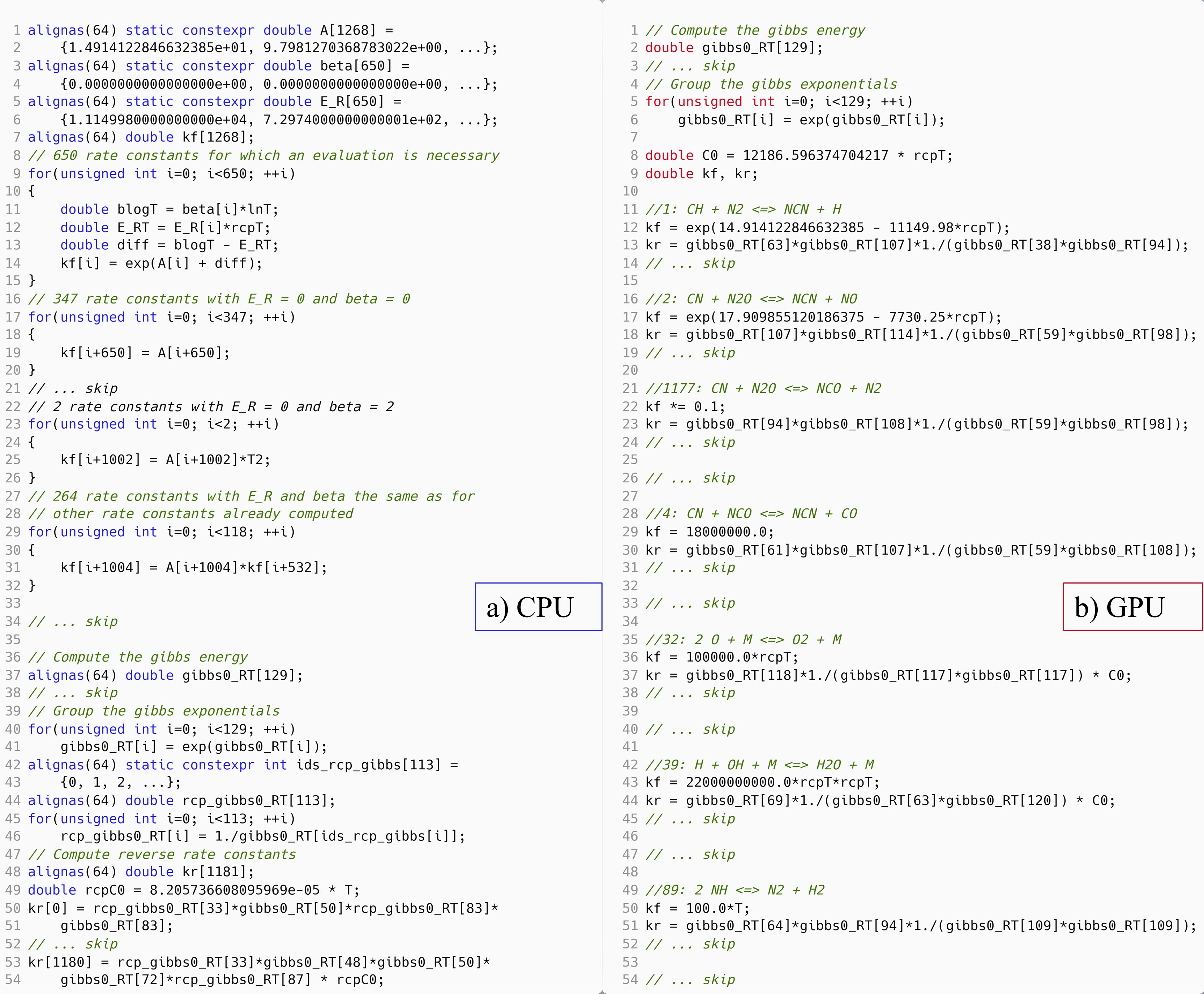}
    }
    \caption{Snippets of the C++ code for the evaluation of the reaction rates.} 
    \label{fig:rates_code_generation} 
    \end{center}
\end{figure}

After eliminating these cases, only 650 of the 1268 exponential functions still have to be evaluated, as highlighted in Fig.~\ref{fig:rates_code_generation}(a) lines 1-32. In addition, all Arrhenius parameters are stored in contiguous arrays and the loop iteration counts are known at compile time, allowing the compiler to vectorize the loops of the grouped rate constants.

The computation of the reverse rate constants $k_r$ requires the evaluation of the equilibrium constants $K_p$ (Eq.~(\ref{eq:K_p})), which involves computing the exponential of the sum of the individual species Gibbs energy $G^{\circ}_k$ for each reaction. For the C$_2$H$_5$OH mechanism, 1180 exponentials would have to be evaluated (the reverse rate constants of irreversible reactions are identically zero). If the exponential of the sum is written as a product of the exponentials of each species, the individual terms can be evaluated using  Eq.~(\ref{eq:Ck}), where all $a_{i,k}, i=0,\cdots 6$ are known parameters from the reaction mechanism and can be stored and aligned in densely packed arrays to enable auto-vectorization. Once $G^{\circ}_k$, the Gibbs energy of the species, has been computed, the exponentials can be evaluated in a vectorized manner (see Fig.~\ref{fig:rates_code_generation}(a) lines 40-41). 
Using this approach, only 129 exponentials are evaluated for $k_r$ in the ethanol mechanism,  compared to 1180 exponentials that would be computed in the generated code presented by Zirwes et al.~\cite{Zirwes2019}. In addition, by pre-computing the reciprocal Gibbs exponentials for the reactants, costly division operations can be replaced with more efficient multiplications in the final calculation for each reaction. Tests on a single AMD EPYC 7763 CPU showed that reducing the number of exponential evaluations for computing reverse rate constants can lead to a 1.5x speedup in the species production kernel for the ethanol mechanism. However, the magnitude of the speedup depends on the mechanism complexity. For instance, with a smaller mechanism like GRI-Mech 3.0~\cite{GRI3.0} with 53 species and 325 reactions the speedup is approximately 1.25x. 

\subsubsection{Mixture-averaged diffusion coefficients} \label{subsec:opt_trans_CPU}

The mixture-averaged diffusivities $D_{km}$ are computed using Eq.~(\ref{eq:mix_diff}), where the $D_{kj}$ binary diffusion coefficients are evaluated using the polynomial fits of Eq.~(\ref{eq:Dkj}). Since only the reciprocal of the binary diffusion coefficients polynomial fits are needed to compute the mixture-averaged diffusivities, the reciprocal polynomial is evaluated directly (Fig.~\ref{fig:diff_code_generation}(a)). In addition, the $D_{kj}$ matrix is symmetric and half of the polynomial evaluations can be avoided. The evaluation of the full $D_{kj}$ matrix can be beneficial in some case since it simplifies the implementation by avoiding the need for special handling of symmetry. Additionally, modern compilers and processors can sometimes better optimize straightforward loops, leading to improved performance due to better cache utilization and fewer branch mispredictions. Furthermore, the full evaluation avoids potential overhead from managing and reusing precomputed values. However, the complete evaluation of the $D_{kj}$ matrix can only be advantageous if the CPU cache is sufficiently large to contain the increased working set of binary diffusion coefficients.

\subsection{Code generation strategy for GPUs}

\subsubsection{Forward and reverse rate constants} \label{subsec:gpu_rates}

In this case, we no longer group the forward rate constants to avoid introducing additional intermediate variables. We can still apply the first two optimizations outlined in Sec.~\ref{subsec:cpu_rates} to effectively reduce the number of exponential evaluations required by Arrhenius law from 1268 to 914 in the C$_2$H$_5$OH mechanism. To address cases where different rate constants share the same values, we reorder the reactions in such a way that reactions with the same $\beta$ and $E_{\mathcal{R}}$ values are grouped together. As shown in Fig.~\ref{fig:rates_code_generation}(b), the $k_f$ from reaction 2 is reused in reaction 1177, which is computed immediately afterwards. This further reduces the exponential evaluations by 264, effectively applying the third optimization described in Sec.~\ref{subsec:cpu_rates} without adding significant register pressure. In the rest of the generated code, each intermediate step leading to the calculation of the progress rates is computed individually for each reaction in order to keep the number of live variables low.

The reverse rates constants are computed as described in Sec.~\ref{subsec:cpu_rates}. 
The transformation of the exponential of the sum to a product does not increase the memory usage since the Gibbs energy array must already reside in memory during the entire computation. 
To optimize memory usage, instead of storing the reciprocals, the divisions are combined into a single fraction,
reducing the number of divisions to only one per reverse rate constant. While GPU division operations are slower, this is more than compensated for by the significant reduction in exponential evaluations.

\subsubsection{Mixture-averaged diffusion coefficients} \label{subsec:opt_trans_GPU}

\begin{figure}[ht]
    \begin{center}
    \makebox[\textwidth]{%
        \includegraphics[width=1.1\linewidth]{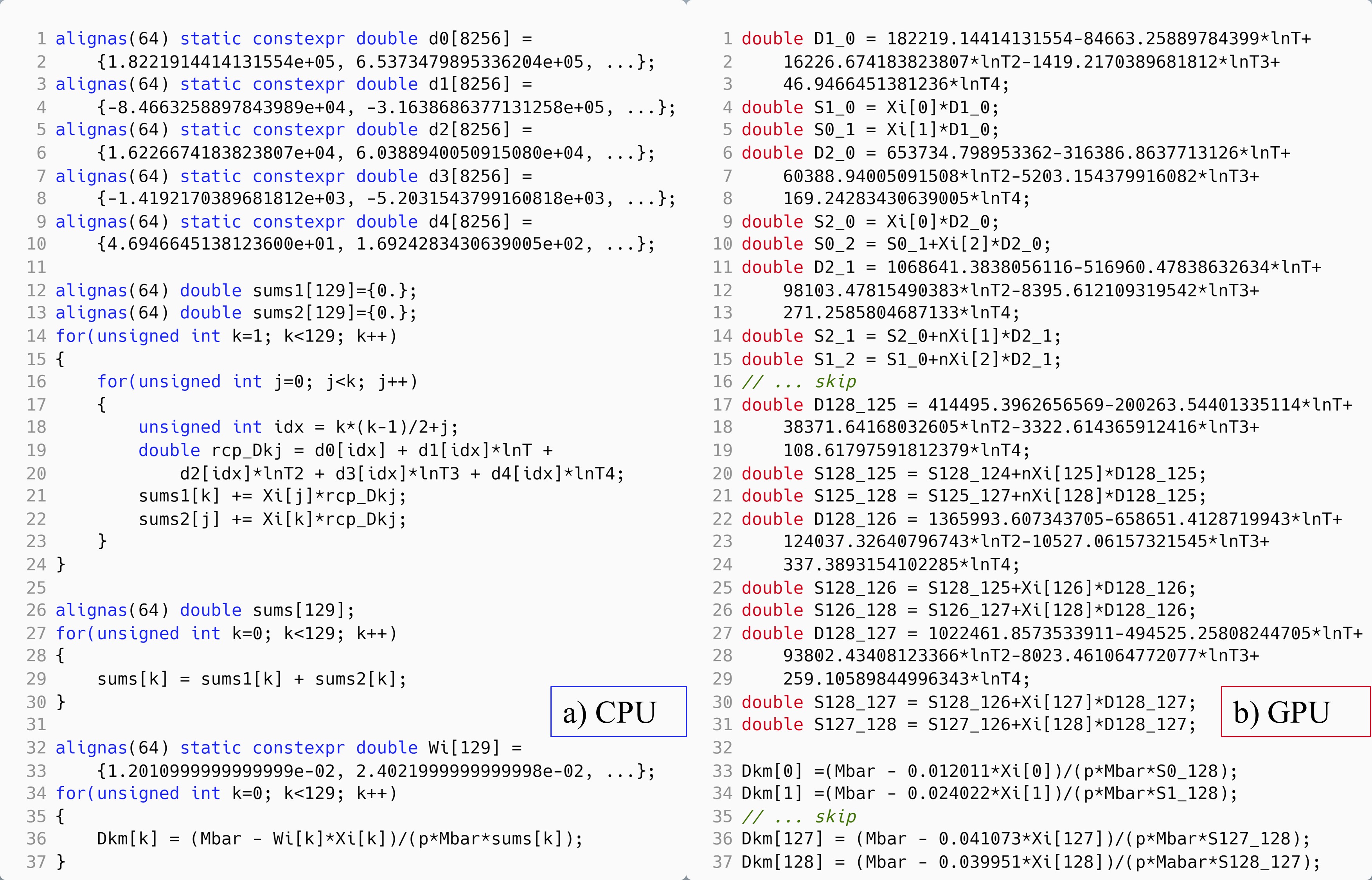}
    }
    \caption{Mixture-average diffusivities code generation.} 
    \label{fig:diff_code_generation} 
    \end{center}
\end{figure}

One way to speedup the computation of the mixture-averaged diffusion coefficients on GPUs is to evaluate the reciprocal polynomial directly (Fig.~\ref{fig:diff_code_generation}(b)) and leverage the symmetry of $D_{kj}$. 
While this can result in significant speed up by minimizing redundant calculations, it also increases the register pressure and can lead to register spilling, low occupancy, and underutilization of mathematical units. Similar to the CPU implementation, another approach can be used that does not assume symmetry and evaluates each entry of the matrix completely. This effectively reduces the number of intermediate values at the cost of doubling the number of polynomial evaluations. The complete evaluation is particularly suitable for large mechanisms since the number of intermediate values that need to be stored in the symmetric evaluation scales quadratically with the number of species.

\subsection{Discussion} \label{subsec:opt_discussion}

We have presented two main approaches for generating code to compute species production rates, thermodynamic and mixture-averaged transport properties designed for CPUs and GPUs, respectively. However, these approaches are not strictly limited to their primary target architectures. The main limitation of applying the first approach to GPUs is the large number of intermediate variables that need to be stored, which scales with the size of the reaction mechanism. Modern server-grade GPUs with their larger cache and increased register files can potentially accommodate these intermediate variables or a significant portion of them for smaller mechanisms. This could lead to higher throughput in some cases when using the first approach on GPUs. Conversely, very large mechanisms (more than 200 species) may generate an excessive number of intermediate values when using the first approach, potentially exceeding the capacity of the CPU cache. This scenario could lead to frequent cache misses and high-latency main memory accesses, so the second approach could be more efficient for CPUs when dealing with such large mechanisms.

In \texttt{KinetiX}, we implemented a grid-level or \textit{per-thread} GPU parallelization model in which each GPU thread independently evaluates the complete species production rates, thermodynamic and mixture-averaged transport properties. The per-thread strategy offers two main advantages: it facilitates the generation of highly optimized code for Single Instruction Multiple Data (SIMD) processors and potentially enables a larger number of concurrent kernel evaluations. However, our current per-thread implementation faces performance limitations due to memory bandwidth constraints, resulting from the limited registers and relatively smaller cache sizes available on GPU streaming multiprocessors (SMs) or compute units (CUs). Alternative parallelization approaches could potentially address these limitations. 
A \textit{per-block} model, where threads within a block collaborate on kernel evaluation, could manage limited registers more effectively by, for example, assigning subsets of reactions to different threads in the species production rates kernel. This could theoretically increase parallelism but can lead to load imbalance due to varying computation times for different reaction types, potentially requiring complex branching or inter-thread synchronization. A \textit{per-warp} model, similar to the Bauer et al. approach~\cite{Bauer2014}, could lead to even better register usage and increased parallelism, but demands advanced branching and inter-warp synchronization techniques. In our current approach the memory requirements scale with the number of species, making it efficient for small- to medium-sized mechanisms depending on GPU architecture. While alternative parallelization approaches can be more efficient for larger mechanisms due to better resource usage, it remains unclear what their effect on overall performance would be given the trade-offs between improved resource utilization and potential overheads from increased synchronization and load balancing complexity. The implications of the chosen parallelization strategy on GPUs and its impact on performance are explored in more detail in the following sections.

\section{Results and discussion} \label{sec:results}

The Python \cite{Python} package \texttt{KinetiX} implements the aforementioned methodology to generate optimized fuel-specific C++ routines to compute chemical kinetics, thermodynamic and mixture-averaged transport properties. The code generator requires the Python modules NumPy \cite{numpy} and ruamel.yaml \cite{ruamel_yaml}, which parses the reaction mechanism file in Cantera YAML format \cite{Cantera} without having to install Cantera. In addition, a Jupyter notebook is used to generate the Cantera reference data used for validation; it requires the Cantera Python package \cite{Cantera} and the SciPy library \cite{SciPy}.

In order to test the correctness and computational performance of the generated kernels, we chose three different reaction mechanisms with increasing size and complexity. Table~\ref{tab:mech} summarizes the chemical kinetics models used as benchmarks in this work, including the H$_2$/O$_2$ model  of Li et al. \cite{Li2004}, the GRI-Mech 3.0 \cite{GRI3.0} model, and the ethanol (C$_2$H$_5$OH) mechanism by Konnov \cite{Konnov2009}.

\begin{table}[H]
\centering
\caption{Combustion mechanisms employed for validation and performance analysis.} \label{tab:mech}
\begin{tabular}{@{}lrr@{}}
\toprule
Mechanism  & \#Species  & \#Reactions    \\
\midrule
H$_2$          & 9          & 21         \\ 
GRI-Mech 3.0   & 53         & 325        \\ 
C$_2$H$_5$OH   & 129        & 1231       \\ 
\bottomrule
\end{tabular}
\end{table}

The performance of \texttt{KinetiX} was evaluated on heterogeneous computing platforms with the hardware specifications reported in Table~\ref{tab:testbeds}. In order to run \texttt{KinetiX} on different computing architectures, the Open Concurrent Compute Abstraction (OCCA) library \cite{OCCA} was used to create kernels that call the automatically generated C++ routines from the Python generator. OCCA is a versatile library that facilitates the development of performance-portable applications. By abstracting the specifics of different parallel programming models, OCCA allows developers to write a single kernel in its language-agnostic format, which can then be compiled and run on multiple hardware backends. This approach ensures that applications can leverage the best performance characteristics of different hardware platforms without the need for extensive rewrites for each target architecture. One of the key features of OCCA, which is the backbone of NekRS \cite{nekRS}, is the ability to perform runtime code generation, where the kernel code is dynamically compiled and optimized for the specific hardware it is running on. In the context of \texttt{KinetiX}, OCCA enables the hybrid MPI+X parallelism approach, where X can be any supported threading model (e.g. CUDA for NVIDIA GPUs, HIP for AMD GPUs, or Data Parallel C++ for Intel GPUs). 

\begin{table}[H]
\centering
\caption{Testbed hardware specifications.} \label{tab:testbeds}
\begin{tabular}{@{}lccc@{}}
\toprule
Processor      & AMD EPYC 9654           & NVIDIA H100 PCIe  & AMD MI250X   \\ 
               & 96@2.4GHz (max 3.7GHz)  & 114SMs@1.8GHz     & 2x110CUs@1.7GHz \\ 
\midrule
Cache          & 384 MB L3  & 50 MB L2          & 16 MB L2              \\ 
Memory         & 1 TB       & 80 GB HBM2e       & 128 GB HBM2e          \\ 
Compiler       & GCC 12.2   & NVCC 12.0         & ROCm 5.2              \\ 
Exec. Space    & Serial     & CUDA              & HIP                   \\ 
\bottomrule
\end{tabular}
\end{table}

The performance of \texttt{KinetiX} was compared to the popular open-source Cantera software package \cite{Cantera} (on CPU only) and CEPTR \cite{PeleSoftware} (on both CPU and GPUs). The similarity of the CEPTR-generated routines facilitated the incorporation of the routines directly into the OCCA kernels of our testing framework, enabling a direct comparison of performance of the \texttt{KinetiX} and CEPTR routines. Further, coupling with the AMReX library \cite{AMReX_JOSS} was required to run the CEPTR-generated code on the different computing platforms presented in Table~\ref{tab:testbeds}. Since Cantera is not designed for GPU execution, performance analysis was limited to the AMD EPYC 9654 CPU.

\subsection{Validation} \label{subsec:validation}

The accuracy of the generated kernels is automatically checked against reference data, which are precomputed using Cantera and a Jupyter notebook that simulates autoignition in a constant pressure reactor. 
Three thermochemical states of pre-ignition, ignition and post-ignition are chosen, and the notebook computes and stores the thermodynamic data together with the reaction rates, the net production rates, and mixture-averaged transport properties. 

Table~\ref{tab:validation} reports the values for the ignition case. The thermodynamic properties and reaction rates show negligible mean and maximum relative differences.
In the mixture-averaged transport properties, the relative differences are moderately higher. The discrepancy was found to be caused by the fitting algorithms used to fit the expressions for the pure species properties. In \texttt{KinetiX} we use the polyfit function of NumPy, while Cantera employs the Eigen library \cite{eigenweb}. The algorithmic differences between the two libraries lead to small differences in the fitted coefficients, which lead to slightly larger errors in the evaluation of the mixture-averaged properties. It is worth mentioning that the relative error between the transport properties of the pure species computed internally in \texttt{KinetiX} and the reference values from Cantera is in general smaller than $10^{-14}$, and the larger errors in the mixture-averaged values are primarily due to the different polynomial fitting functions. Nevertheless, the maximum relative errors for the mixture-averaged transport properties are below $10^{-7}$.

\begin{table}[H]
\centering
\caption{Summary of difference between \texttt{KinetiX}-generated properties and Cantera reference values. Error statistics are based on the mean and maximum relative error $E_{rel}$ for each property.}
\label{tab:validation}
\begin{tabular}{lccc}
\toprule
Model & Properties & Mean & Maximum \\
\midrule
\multirow{3}{*}{H$_2$} & Thermodynamic properties & 1.987 × 10$^{-16}$ & 4.220 × 10$^{-16}$ \\
 & Reaction rates & 1.447 × 10$^{-11}$ & 8.677 × 10$^{-11}$ \\
 & Transport properties & 3.765 × 10$^{-10}$ & 1.800 × 10$^{-8}$ \\
\midrule
\multirow{3}{*}{GRI-Mech 3.0} & Thermodynamic properties & 2.623 × 10$^{-16}$ & 7.681 × 10$^{-16}$ \\
 & Reaction rates & 6.160 × 10$^{-13}$ & 1.899 × 10$^{-11}$ \\
 & Transport properties & 1.365 × 10$^{-9}$ & 1.126 × 10$^{-8}$ \\
\midrule
\multirow{3}{*}{C$_2$H$_5$OH} & Thermodynamic properties & 7.292 × 10$^{-17}$ & 6.015 × 10$^{-16}$ \\
 & Reaction rates & 3.629 × 10$^{-11}$ & 2.329 × 10$^{-9}$ \\
 & Transport properties & 2.826 × 10$^{-10}$ & 9.874 × 10$^{-8}$ \\
\bottomrule
\end{tabular}
\end{table}

\subsection{Performance analysis} \label{subsec:performance}

The performance of the \texttt{KinetiX}-generated routines was tested by evaluating the species production rates and the transport kernels for the three reaction mechanisms on the three testbeds described in Tables \ref{tab:mech} and \ref{tab:testbeds}, respectively. The thermodynamic kernel was excluded because it only involves basic polynomial evaluations. The input to the kernels consists of a vector containing species mass fractions and temperature, multiplied by the number of individual states, corresponding to simulated grid points, to produce a comprehensive vector of thermochemical composition states. Performance metrics are based on throughput, defined as the number of reactions evaluated per second for the species production rates kernel (measured in giga-reactions per second, GRXN/s) and the number of mixture-averaged transport properties computed in the transport kernel (measured in giga-degrees-of-freedom per second, GDOF/s). The absolute times for each experiment can be computed by dividing the number of reactions or degrees of freedom, respectively, by the reported throughput defined as the arithmetic mean of 50 repetitions of each experiment. 

For comparison, we evaluated the performance against both CEPTR and Cantera using the same metrics. While a direct comparison of species production rates was possible since all three packages implement the same formulations for chemical kinetics (detailed in \ref{sec:theory}), the comparison of transport properties required special consideration. CEPTR follows the CHEMKIN approach by employing third-degree polynomials and evaluating the logarithm of the transport properties, whereas both \texttt{KinetiX} and Cantera use fourth-degree polynomials and evaluate the properties directly. In addition, CEPTR only generates the fitting coefficients for the transport quantities, with mixture-averaged properties evaluated internally. These differences in transport property computation precluded a direct performance comparison between \texttt{KinetiX} and CEPTR for the transport kernel, although we were able to compare CPU performance with Cantera due to their similar approaches.


\subsubsection{Species production rates}

\begin{figure}[ht]
	\begin{center}
	\includegraphics*[width=0.8\linewidth]{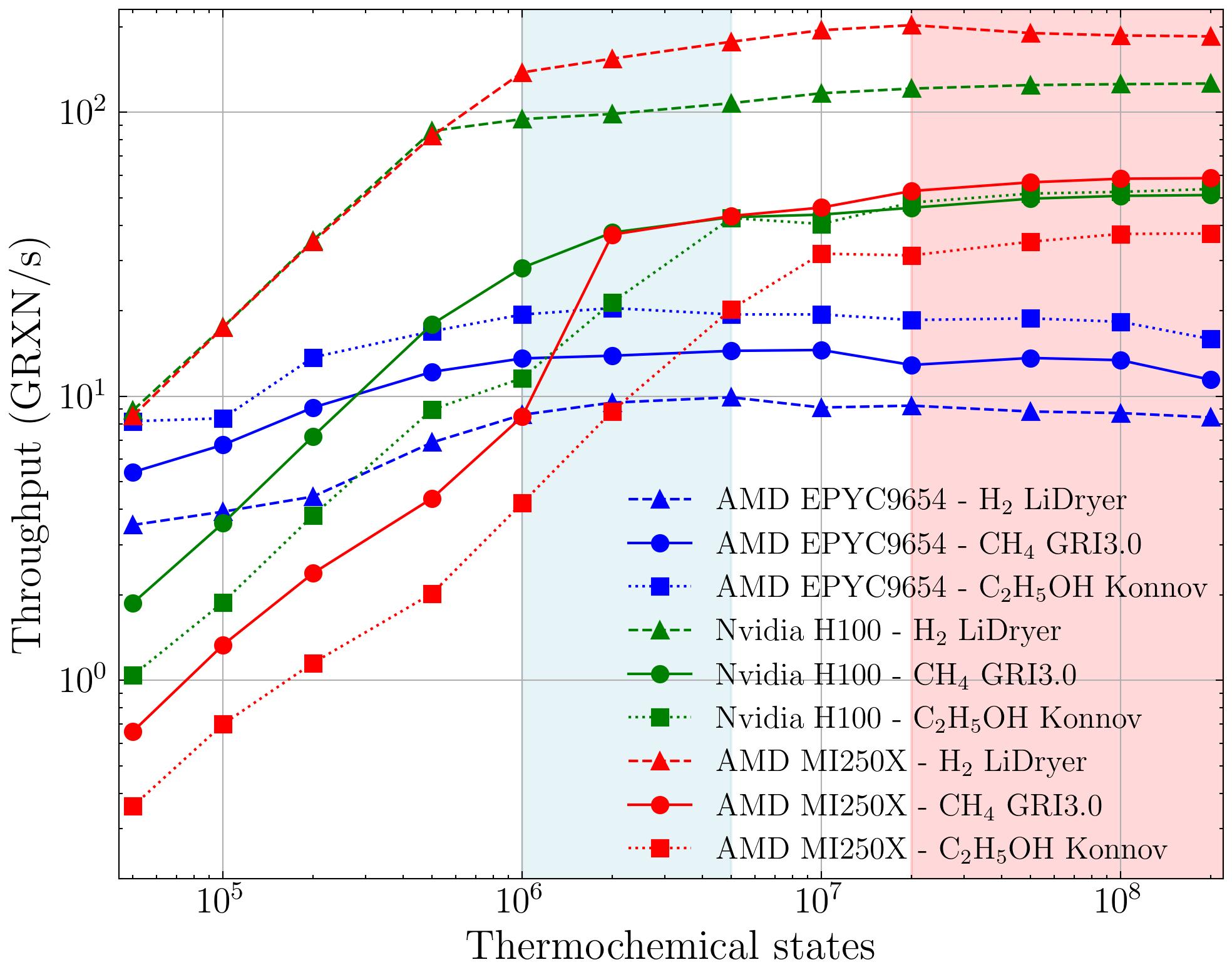}
	\caption{Throughput (in GRXN/s) versus number of thermochemical states for \texttt{KinetiX} species production rates kernel on different platforms. The blue and red shaded areas mark the ranges of peak throughput.} 
  \label{fig:throughput_rates} 
	\end{center}
\end{figure}

Figure~\ref{fig:throughput_rates} compares the computed throughput as a function of the thermochemical states, where an increase in the number of states mimics an increase in the problem size. For the GPU implementations (blue and red curves), all mechanisms show an initial increase in throughput as the number of states increases. This trend continues until the GPU is fully utilized, and the throughput growth rate begins to plateau. Notably, this point of thread saturation occurs at approximately the same number of conditions for each reaction model due to the chosen parallelization approach. 

In contrast, CPU performance exhibits higher initial throughput values and reaches its peak performance at almost two orders of magnitude smaller number of thermochemical states than the GPUs. The earlier saturation point underscores the CPU's ability to achieve optimal performance on smaller problem sizes. These observations highlight a crucial consideration when evaluating the performance of such kernels across different computing platforms, namely that there is an optimal performance window with respect to problem size (blue and red shaded areas in Fig.~\ref{fig:throughput_rates}) for which peak throughput is achieved. While GPUs generally offer superior performance for large-scale problems, CPUs can actually achieve higher throughput for smaller problem sizes.

\begin{figure}[ht]
	\begin{center}
	\includegraphics*[width=\linewidth]{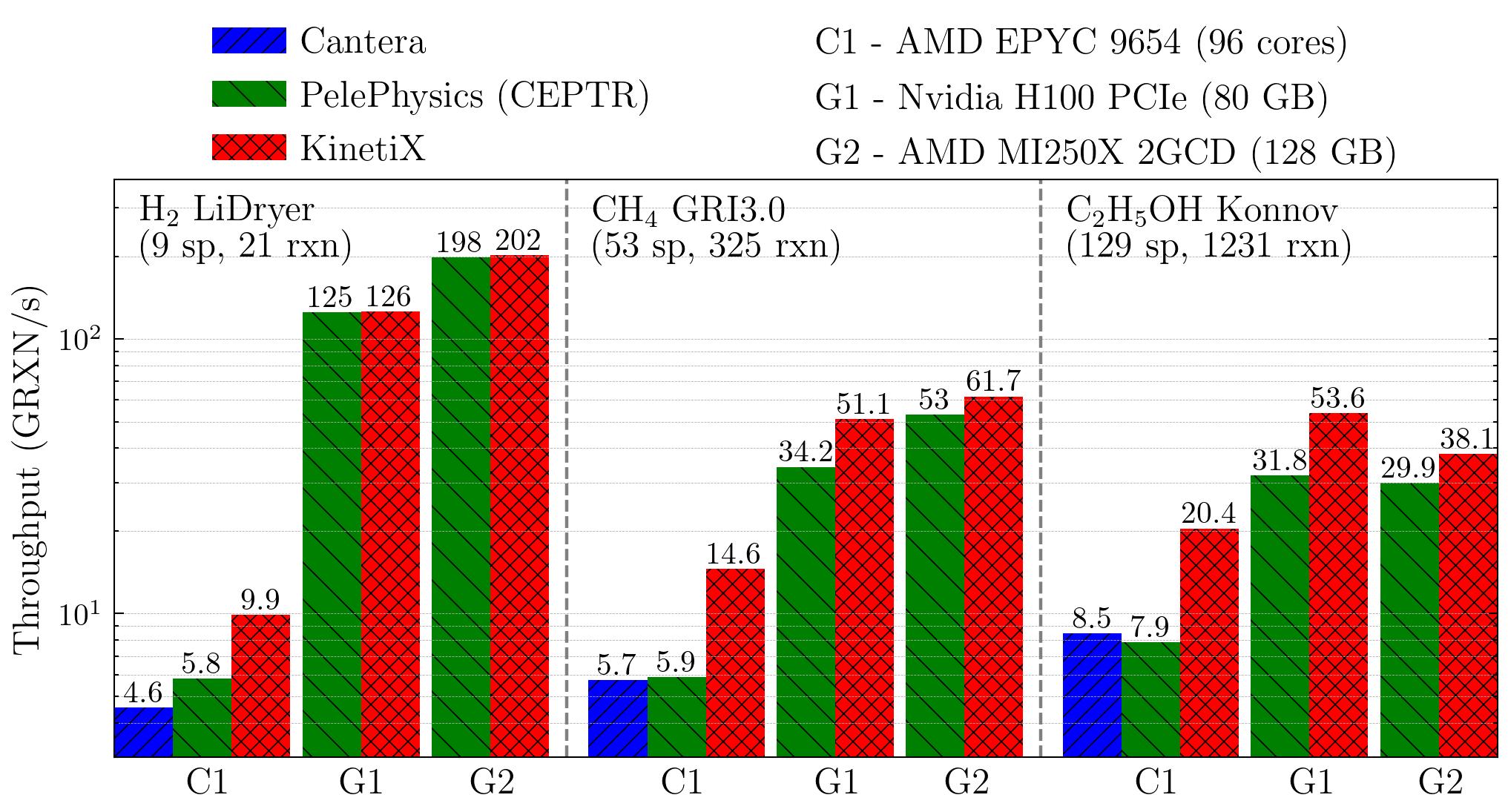}
	\caption{Comparison of peak throughput (measured in GRXN/s) between \texttt{KinetiX} (red bars), CEPTR (green bars) and Cantera (blue bars) for the three kinetic models on the three computing platforms.} 
  \label{fig:performance_comparison_rates} 
	\end{center}
\end{figure}

Figure~\ref{fig:performance_comparison_rates} presents a comparative analysis of the production rates kernel performance between \texttt{KinetiX}, CEPTR, and Cantera. The throughput is evaluated as the peak value obtained within the optimal performance window. It can be observed that the \texttt{KinetiX} performance on GPUs (red bars) decreases with increasing mechanism size. This is due to the adopted \textit{per-thread} GPU parallelization model (Sec.~\ref{subsec:opt_discussion}.) While this model can theoretically allow for more concurrent kernel evaluations, it faces performance limitations with larger reaction mechanisms, due to increasing memory requirements per thread, which scales with the number of species. 
Although GPUs can hide the long-latency operations in the computation of the reaction rates through massive parallelism, this parallelism is limited by register usage. As the size of mechanisms increases, the increased register pressure leads to lower concurrency, as fewer threads can be simultaneously active on each SM or CU, and register spilling, where data spills into slower memory hierarchies. Both effects introduce additional latencies and reduce the GPU ability to hide memory and instruction latencies, thereby decreasing overall throughput. These limitations become more apparent for larger reaction mechanisms, which explains why the highest throughput is achieved for the smallest mechanism, H$_2$, where register usage remains low enough to maintain high concurrency and avoid spilling.

When comparing the NVIDIA H100 and AMD MI250X GPUs, a relatively large difference in throughput in favor of the MI250X can be observed for the H$_2$ mechanism. In this case, up to 46 double precision variables remain live throughout the kernel execution, requiring 92 32-bit registers. Since both the H100 and MI250X GPUs offer up to 255 registers per thread, each capable of storing 32 bits of information, the H$_2$ mechanism can be easily accommodated within the available registers per thread on both GPUs, avoiding register spilling. This allows the GPUs to utilize their threads effectively, and the MI250X's Dual Graphics Compute Die (GCD) design with larger register file sizes and more CUs give it a theoretical advantage in concurrent thread execution. In addition, the MI250X is optimized for high-performance computing applications and excels in double precision (FP64) computations. In contrast, the H100 is optimized for a broader range of precision levels, including FP64, FP32, FP16 and FP8, to cater to diverse computational tasks. These two factors can explain the significant throughput difference for the H$_2$ mechanism. However, when comparing the throughput for FP32, the H100 demonstrates substantial improvements, reaching up to 242 GRXN/s compared to the 274 GRXN/s of the MI250X in the single precision evaluation of the production rates kernel. Nevertheless, since the accuracy of the production rates kernel is crucial in combustion simulations, it is expected that double precision values will be used most of the time, and therefore a more in-depth comparison of single precision performance on the different architectures is not presented here. 

For the GRI 3.0 mechanism, there is a significant decrease in throughput on both GPUs, which is due to the significantly higher register pressure resulting from the 178 live double-precision variables, i.e. 1376 bytes of information that needs to be stored. Even with the maximum number of registers available per thread (255), 404 bytes of data are spilled to off-chip global memory. In addition, this high register pressure leads to lower occupancy and under-utilization of mathematical units. The NVIDIA H100 GPU features a large L2 cache of 50 MB. While spilled data is initially stored in global memory, the frequent accesses to this data can lead to it being cached in the L2, which can significantly reduce latency compared to repeatedly fetching the data directly from global memory. The large cache size increases the likelihood that frequently accessed spilled data remains in cache, potentially offering substantial performance benefits. In contrast, the AMD MI250X, with an L2 cache size of 16 MB, presents more of a challenge. The limited cache memory exacerbates the effects of the data spill, significantly decreasing throughput. The difference in performance between the two GPUs is also significantly smaller for the CH$_4$ mechanism.

The largest mechanism tested for ethanol requires 401 double-precision variables and 3208 bytes of information per thread. With the maximum number of registers per thread fully utilized, this still results in 2288 bytes of spilled information. For the MI250X, its relatively smaller L2 cache means that a significant amount of data may frequently need to be fetched from global memory. This can result in more frequent slow global memory accesses, impacting performance. In contrast, the H100's larger L2 cache can cache more of the frequently accessed spilled data. While the initial spill to global memory adds some latency, subsequent accesses to this data, if cached, would experience significantly less latency compared to repeatedly fetching from off-chip global memory, thus achieving higher throughput.

When comparing the performance between \texttt{KinetiX} and CEPTR on GPUs (red and green bars, respectively, in Fig.~\ref{fig:performance_comparison_rates}), we observe that \texttt{KinetiX} consistently performs better for all mechanisms. Both codes utilize an approach where reactions are unrolled and computed sequentially on GPUs with similar number of live variables during kernel execution. The primary distinction between the two codes lies in the reduction of exponential evaluations described in Sec.~\ref{sec:code_generation}. By pre-computing the Gibbs exponentials, \texttt{KinetiX} significantly reduces the number of evaluations required to compute the reverse rate constants, which scales with the ratio of the number of reactions to the number of species. This reduction becomes increasingly beneficial for larger mechanisms. Similarly, the reduction of exponential evaluations in Arrhenius law (Sec.~\ref{sec:code_generation}) is proportional to the mechanism size, making the \texttt{KinetiX} optimization strategy more effective with increasing mechanism size. For hydrogen, the performance difference is small, with both codes achieving similar throughput. The H$_2$ mechanism, with 9 species and 21 reactions, allows for only 12 avoided exponential evaluations at the cost of 21 additional divisions, which negatively impact performance. Moreover, only two reactions benefit from reduced exponentials in Arrhenius law. Consequently, the overall performance difference between \texttt{KinetiX} and CEPTR remains small on both GPUs. For methane, the performance difference increases, with \texttt{KinetiX} achieving a speedup of up to 1.49x on the H100 GPU and 1.16x on the MI250X GPU. This improvement is due to the mechanism having a higher reaction-to-species ratio and a larger number of exponentials falling into the strategies described in Sec.~\ref{sec:code_generation}. The ethanol mechanism reinforces this trend, with the performance benefit of \texttt{KinetiX} increasing further to 1.69x on the H100 GPU and 1.27x on the MI250X GPU.

On the CPU, throughput values increase with growing mechanism size, contrary to GPU performance. This is primarily due to the larger cache size offered by CPUs like the AMD EPYC 9654, which provides up to 384 MB of L3 cache. This ample cache allows even the largest mechanisms to be fully stored in fast memory, enabling efficient on-chip computation and memory bandwidth utilization. \texttt{KinetiX} performance significantly improves with increasing mechanism size due to the optimization approaches described in Sec.~\ref{sec:code_generation}. These changes maximize data cache usage and auto-vectorization through cache-friendly data structures and linear data access patterns, while also reducing the number of exponential evaluations and redundant operations. Consequently, \texttt{KinetiX} achieves up to 2.40x higher throughput compared to Cantera and 2.58x higher compared to CEPTR on the CPU for the largest tested mechanism for C$_2$H$_5$OH.
CEPTR and Cantera also show performance improvements with increasing mechanism size, albeit at a lower rate. This can be attributed to the presence of more reactions with $E_R=\beta=0$, which are treated similarly as in \texttt{KinetiX}. Additionally, the ratio of standard reversible reactions to more complex falloff reactions increases with the mechanism size, leading to a decrease in the average time required to compute a reaction and consequently to higher throughput.

\subsubsection{Mixture-averaged transport properties}

\begin{figure}[ht]
	\begin{center}
	\includegraphics*[width=0.8\linewidth]{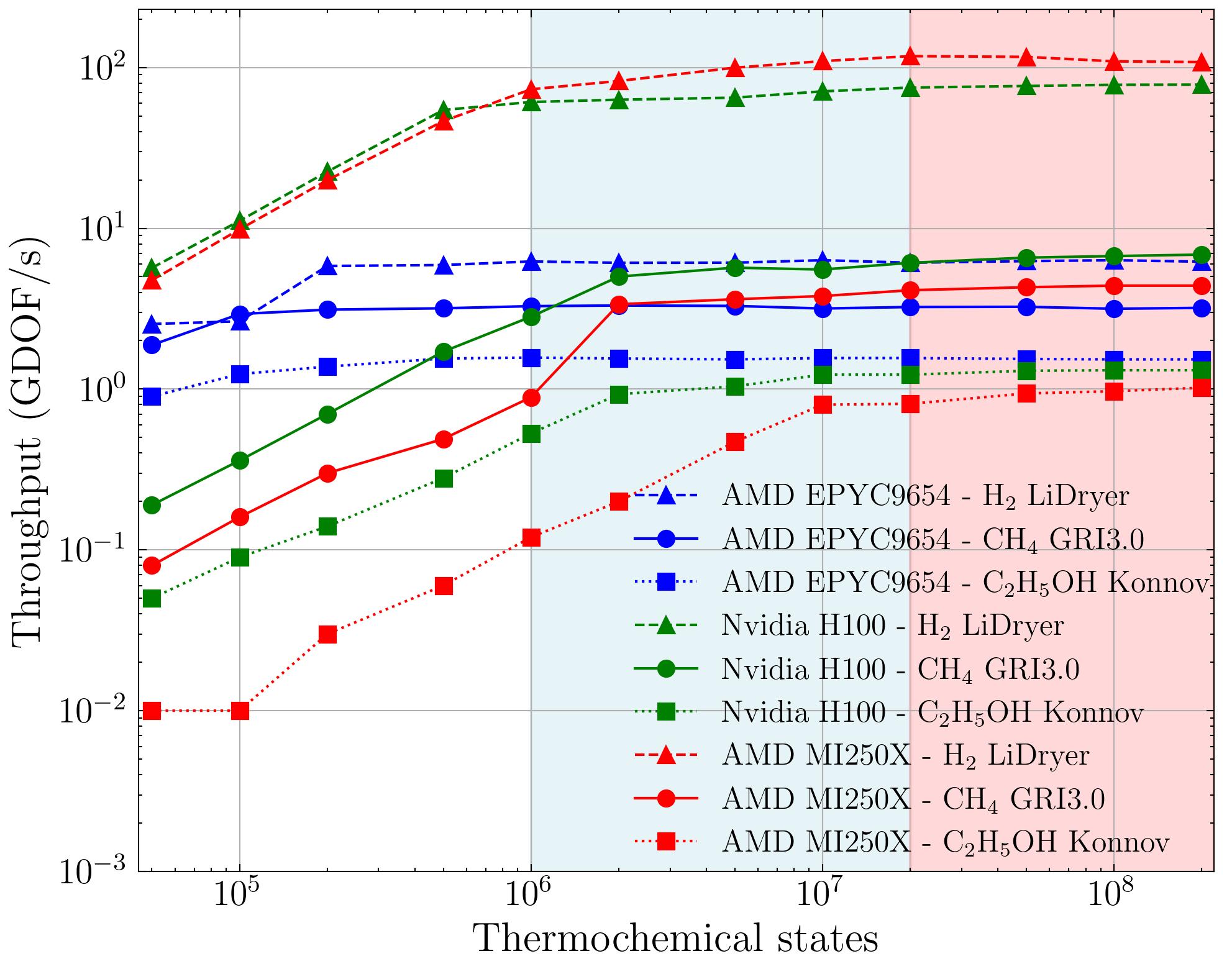}
	\caption{Throughput (in GDOF/s) versus number of thermochemical states for the \texttt{KinetiX} transport kernel. The blue and red shaded areas indicate ranges of peak throughput.} 
  \label{fig:throughput_transport} 
	\end{center}
\end{figure}

Figure~\ref{fig:throughput_transport} depicts the computed throughput of the transport properties kernel as a function of the number of thermochemical states. The GPU implementations show a pattern similar to that observed in the production rates kernel. Initially, the throughput increases steadily as the number of states grows, continuing until the GPU reaches full utilization. Again, this saturation point occurs at a significantly higher number of states than for the CPU implementation. In contrast, the CPU shows much higher initial throughput values and reaches peak performance at a much lower number of thermochemical states than the GPU. After reaching this peak, throughput remains approximately constant over a wide range of thermochemical states. This sustained performance is due to the AMD EPYC 9654 sizable L3 cache size (384MB), which allows the CPU to maintain peak performance even as the problem size increases beyond the initial saturation point. It should be noted that for a CPU with a smaller cache, the throughput curve would likely exhibit more of a bell shape. In such cases, performance would decrease when the problem size becomes too large to fully fit in the cache. The performance degradation occurs as the processor becomes increasingly reliant on slower main memory accesses, resulting in a decrease in throughput for a larger number of thermochemical states.

\begin{figure}[ht]
	\begin{center}
	\includegraphics*[width=\linewidth]{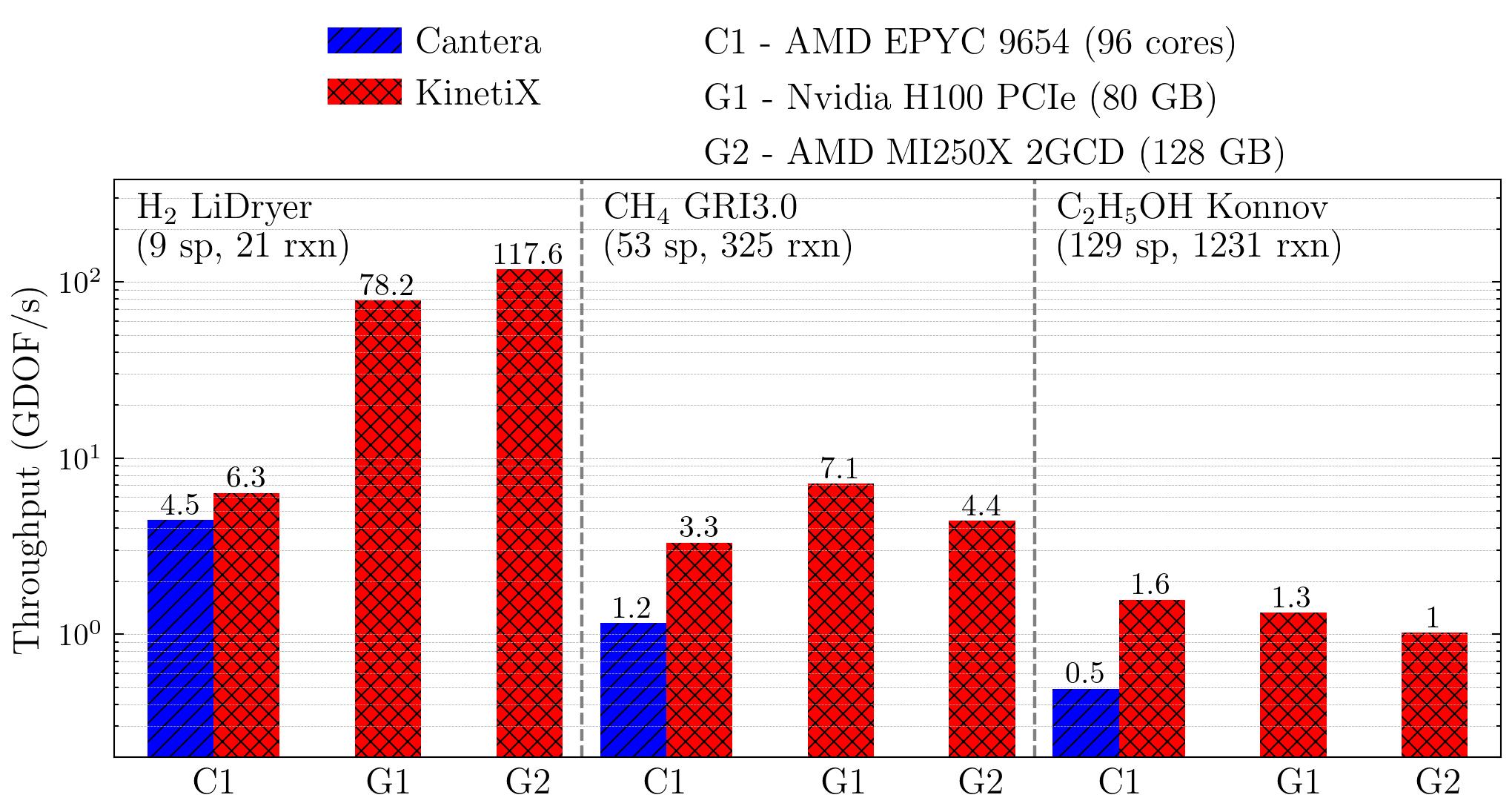}
	\caption{Comparison of peak throughput (measured in GDOF/s) between \texttt{KinetiX} (red bars) and Cantera (blue bars) for the three kinetic models on the three computing platforms.} 
  \label{fig:performance_comparison_transport} 
	\end{center}
\end{figure}

Figure \ref{fig:performance_comparison_transport} shows the maximum achieved throughput in GDOF/s for \texttt{KinetiX} on GPUs and CPUs and Cantera on CPUs. The transport kernels on GPUs exhibit behavior similar to that of the species production rates kernels. For hydrogen, the throughput is high, with MI250X achieving higher values compared to H100. In the transport kernel, the number of live variables can scale quadratically with the number of species, often exceeding the small on-chip memory allocated to each thread, even for small mechanisms, resulting in register spilling, and low occupancy. However, by using techniques such as loop unrolling, computing the contributions of individual species to the mixture-averaged viscosity, and evaluating the complete diffusivities matrix (Sec.~\ref{subsec:opt_trans_GPU}), the number of live variables in the transport kernels can be significantly reduced. As a result, these optimized transport kernels require fewer live variables compared to the species production rates kernel. For hydrogen, the memory requirements per thread can easily be accommodated within the limit of 255 registers per thread of each GPU, at a value of 50 registers per thread. The lower number of registers results in higher occupancy compared to the production rates kernel. The larger register file size of the MI250X allows more threads per CU to remain active, which in combination with the higher number of CUs, leads to greater theoretical parallelism for the H$_2$ mechanism and higher throughput. For the GRI-Mech 3.0 mechanism, the larger number of live variables results in higher register pressure, lower occupancy and data  spillage to off-chip memory. The H100's increased L2 cache size, which is more than three times the size of the MI250X, allows it to manage this spilled data more effectively, resulting in better performance. For the C$_2$H$_5$OH mechanism, the transport kernel becomes global memory-bound on both GPU architectures, resulting in significantly lower throughput.

It should be pointed out that unlike the GRXN/s metric used for species production rates, the GDOF/s metric used for mixture-averaged transport properties is not independent of the problem size. This is because the number of floating point operations for transport properties scales quadratically with the number of species. Consequently, the throughput values decrease with increasing mechanism size, even on CPUs. This effect is observed despite the large cache size of the AMD EPYC 9654 CPU, which can accommodate even the largest mechanism.
In comparison to Cantera, the throughput improvement becomes more pronounced with larger mechanism sizes. For mixture-averaged transport properties, the coefficients for the polynomial fits of the pure species properties are densely packed in memory and defined as compile-time constants, which together with loop optimization strategies enables the auto-vectorization of the loops for a more efficient computation of the mixture-averaged transport properties.
For the largest mechanism, a significant speedup of 3.2x is achieved compared to Cantera. This demonstrates the effectiveness of our optimization techniques on CPUs, particularly for complex combustion kinetics.

\section{Conclusions} \label{sec:conclusions}

We present \texttt{KinetiX}, a software toolkit designed to generate species production rates, thermodynamic and mixture-averaged transport properties routines for high-performance execution on both CPU and GPU architectures. The code generator parses a Cantera YAML file containing the reaction mechanism data to generate fuel-specific source code files tailored to each architecture. 

The generated C++ routines for CPUs optimize chemical rates computations by restructuring and reordering the reaction mechanism and eliminating redundant operations. Furthermore, data alignment and loops with trivial access patterns enable auto-vectorization, reducing the latency of costly mathematical instructions. For mixture-averaged transport properties, densely packed polynomial coefficients and compile-time constants further improve performance. On GPUs, \texttt{KinetiX} enhances performance by manually unrolling loops, reducing the number of expensive exponential evaluations and keeping the number of live variables low for better register usage. This approach keeps more threads active, which helps to better hide the long-latency mathematical instructions. In contrast to the CPU approach, rate constants and reactions are not grouped together to avoid storing intermediate values; instead, the progress rates are computed individually for each reaction. In addition, loops for mixture-averaged transport properties are unrolled to avoid large working sets. The accuracy of the \texttt{KinetiX}-generated routines is validated against Cantera reference values. 

In order to run \texttt{KinetiX} on different computing architectures, the OCCA library, which allows for runtime code generation for different threading programming paradigms was used to create kernels that call the generated C++ routines. The kernel performance was evaluated on some of the latest CPU and GPU architectures from AMD and NVIDIA i.e., AMD EPYC 9653, AMD MI250X and NVIDIA H100. Performance benchmarking showed that the routines generated by \texttt{KinetiX} outperform the general-purpose Cantera library with speedups of up to 2.4x for species production rates and 3.2x for mixture-averaged transport properties on CPU. Compared to CEPTR, \texttt{KinetiX} achieves speedups of up to 2.6x on CPU and up to 1.7x on GPUs for the species production rates kernel on a single-threaded basis. The most significant performance improvements were observed with the largest reaction mechanism (C$_2$H$_5$OH), whereas the smallest mechanism tested (H$_2$) yielded more modest gains.

Planned development work for \texttt{KinetiX} includes support for quasi-steady-state approximation (QSSA) species and analytical Jacobian computation. Further studies will focus on exploring different parallelization strategies for GPUs. Currently, the \textit{per-thread} or grid-based parallelization approach used in \texttt{KinetiX} is effective for small to medium sized kinetic problems. However, as HPC trends move towards GPUs with larger register count and cache size, these limitations may be mitigated in the future, enabling the efficient computation of larger mechanisms. Furthermore, the reaction mechanisms for carbon-free fuels generally involve fewer species and reactions. Nonetheless, exploring alternative strategies, such as collaborative thread execution or warp specialization techniques, would be valuable to determine the best performance scaling with reaction mechanism size on GPUs. In addition, we will explore mixed precision formulations for the generated routines which can potentially offer significant performance on GPUs.

\section*{Acknowledgements}
The authors gratefully acknowledge the Gauss Centre for Supercomputing e.V. (\url{www.gauss-centre.eu}) for funding this project by providing computing time on the GCS Supercomputer JUWELS at Julich Supercomputing Centre (JSC). The authors gratefully acknowledge the EuroHPC Joint Undertaking for awarding this project access to the EuroHPC supercomputer LUMI, hosted by CSC (Finland) and the LUMI consortium through a EuroHPC Early Access call.

\section*{Funding} 
This project received funding from the European Union’s Horizon 2020 research and innovation program under the Center of Excellence in Combustion (CoEC) project, grant agreement No 952181. 

\section*{Conflict of Interest} 
The authors declare that they have no conflict of interest.
\section*{CRediT authorship contribution statement} \textbf{Bogdan A. Danciu}: Conceptualization, Methodology, Software, Validation, Formal analysis, Data curation, Visualization, Writing - original draft, Writing - review and editing. \textbf{Christos E. Frouzakis}: Conceptualization, Funding acquisition, Resources, Project administration, Writing - review and editing. 

\section*{Data Availability} 
The data are available from the authors upon reasonable request.

\appendix

\section{Theoretical background} \label{sec:theory}

For completeness, the appendix summarizes the expressions used in the generator for the species production rates, thermodynamic and transport properties. More comprehensive presentations can be found in ~\cite{warnatz2006, law2010, glassman2014, Kee2017}.

\subsection{Thermodynamic properties}

The standard-state thermodynamic properties , namely the molar heat capacity at constant pressure  $C^\circ_{p,k}$, molar enthalpy $H^\circ_k$, and molar entropy $S^\circ_k$ for a gaseous species $k$ are given in terms of seven-coefficient polynomial fits \cite{gordon1994}:

\begin{eqnarray}
    \frac{C^\circ_{p,k}}{\mathcal{R}} &=& a_{0,k} + a_{1,k}T + a_{2,k}T^2 + a_{3,k}T^3+ a_{4,k}T^4,  \\
    \frac{H^\circ_k}{\mathcal{R}T} &=& a_{0,k} + \frac{a_{1,k}}{2}T + \frac{a_{2,k}}{3}T^2 + \frac{a_{3,k}}{4}T^3+ \frac{a_4,k}{5}T^4 + \frac{a_{5,k}}{T}, \label{eq:H_RT}\\
    \frac{S^\circ_k}{\mathcal{R}}  &=&  a_{0,k}\ln T + a_{1,k}T + \frac{a_{2,k}}{2}T^2 + \frac{a_{3,k}}{3}T^3+ \frac{a_4,k}{4}T^4 + a_{6,k}, \label{eq:S_R}
\end{eqnarray}
where $T$ is  temperature and $a_{i,k}, \,\, i=0,\cdots 6$ are the polynomial coefficients for species $k$, and $^\circ$ refers to the standard state at one atmosphere; for calorically-perfect gasses, the standard-state values are pressure independent. 

If the thermodynamic properties in molar units are divided by the species molecular weight $W_k$ the mass-related properties are obtained. The specific heat and enthalpy (in mass units) are therefore defined as: 
\begin{eqnarray}
    c_{p,k} &=& \frac{C_{p,k}}{W_k} \quad \text{and} \quad h_{p,k} = \frac{H_{p,k}}{W_k}.
\end{eqnarray}

The mixture-average specific heat at constant pressure is 
\begin{eqnarray}
    c_p &=& \sum_{k=1}^{N_s} Y_k c_{p,k}.
\end{eqnarray}

\subsection{Chemical kinetics}

For a reaction mechanism with $N_r$ gas-phase reactions between $N_s$ chemical species, the net production rate of species $k$ is
\begin{equation}
    \dot{\omega}_k = \sum_{i=1}^{N_r} \nu_{ki} R_i,
\end{equation}
with $\nu_{ki}=\nu''_{ki}-\nu'_{ki}$ being the net stoichiometric coefficient of species $k$ in reaction $i$, and $R_i$ the rate of progress of reaction $i$
\begin{equation}
    R_i = c_i r_i,
\end{equation}  
with $r_i$ being the production rate of reaction $i$, and  $c_i$ the third-body/pressure modification given by
\begin{equation}
    c_i =
    \begin{cases}
        \begin{aligned}
            &1 \quad &&\text{for elementary reactions} \\
            &[X]_i \quad &&\text{for third-body reactions}  \\
            &\frac{P_{r,i}}{1+P_{r,i}} F_i \quad &&\text{for unimolecular/recombination falloff reactions} \\
            &\frac{1}{1+P_{r,i}} F_i \quad &&\text{for chemically-activated bimolecular reactions} \label{eq:c_i}
        \end{aligned}
    \end{cases}
\end{equation}
The third-body concentration for the $i$-th reaction $[X]_i$, reduced pressure $P_{r,i}$, and falloff blending factor $F_i$ is defined in the following sections.

The production rate of the $i$-th elementary reaction is defined as 
\begin{eqnarray}
    r_i &=& k_{f,i}R_{f,i} - k_{r,i}R_{r,i} \\
    r_i &=& k_{f,i}\prod_{k=1}^{N_s} [X_k]^{\nu'_{ki}} -
          k_{r,i}\prod_{k=1}^{N_s} [X_k]^{\nu''_{ki}}, \label{eq:r_i}
\end{eqnarray}
where $k_{f,i}, k_{r,i}$ are the rate constants of the forward and reverse reaction, and $\nu'_{ji}, \nu''_{ji}$ the stoichiometric coefficients of the participating reactants and products, respectively, and $[X_k]$ denotes molar concentration.

The forward rate constant for the $i$-th reaction follows the modified Arrhenius law:
\begin{equation}
    k_{f,i} = A_iT^{\beta_i} \exp\left( \frac{-E_{{\mathcal{R}}, i}}{T} \right), \label{eq:arrhenius}
\end{equation}
where $A_i$ is the pre-exponential factor, $\beta_i$ the temperature exponent and $E_{\mathcal{R},i}$ = $E_{a,i}/\mathcal{R}$ the activation temperature;  $\mathcal{R}$ is the ideal gas constant.

Depending on the Arrhenius parameters, the computational cost for the calculation of the forward rate constant can be reduced \cite{law2010}. The effects of these formulations are further discussed in Sec.~\ref{sec:code_generation}.
\begin{equation}
    k_{f,i} =
    \begin{cases}
        \begin{aligned}
            &\exp(\log A_i + \beta_i\log T - E_{\mathcal{R},i}/T) \quad &&\text{if $\beta_i \ne 0$ and $E_{\mathcal{R},i} \ne 0$,} \\
            &\exp(\log A_i + \beta_i\log T) \quad &&\text{if $\beta_i\ne0$ and $E_{\mathcal{R},i}=0$,}  \\
            &\exp(\log A_i - E_{\mathcal{R},i}/T) \quad &&\text{if $\beta_i=0$ and $E_{\mathcal{R},i} \ne 0$,} \\
            &A_i \quad &&\text{if $\beta_i=0$ and $E_{\mathcal{R},i}=0$,} \\
            &A_i \prod_{}^{\beta_i}T \quad &&\text{if $E_{\mathcal{R},i}=0$ and $\beta_i \in \mathbb{Z}$,} \label{eq:k_f}
        \end{aligned}
    \end{cases}
\end{equation}
where $\mathbb{Z}$ is the set of integer numbers. 

For reversible reactions, the reverse rate constants $k_{r,i}$ are related to the forward rate constants through the equilibrium constants 
\begin{equation}
    k_{r,i} = \frac{k_{f,i}}{K_{c,i}}.
\end{equation}
The equilibrium constants can be calculated more conveniently from the thermodynamic properties with respect to pressure, although $K_{c,i}$ is expressed with respect to concentration. These two quantities are connected by
\begin{equation}
    K_{c,i} = K_{p,i} \left(\frac{p_{\text{atm}}}{\mathcal{R}T}\right)^{\sum_{k=1}^{Ns} \nu_{k,i}},
\end{equation}
where $p_{\text{atm}}$ is the pressure corresponding to one standard atmosphere.

The equilibrium constants $K_{p,i}$ are given by 
\begin{equation}
    K_{p,i} = \exp \left( \frac{\Delta S^\circ_i}{\mathcal{R}} - \frac{\Delta H^\circ_i}{\mathcal{R}T} \right).
\end{equation}
$\Delta$ denotes the difference that occurs in the complete transition from reactants to products in the $i$-th reaction
\begin{eqnarray}
    \frac{\Delta S^\circ_i}{\mathcal{R}} &=& \sum_{k=1}^{N_s}\nu_{k,i}\frac{S^\circ_k}{\mathcal{R}}, \\
    \frac{\Delta H^\circ_i}{\mathcal{R}T} &=& \sum_{k=1}^{N_s}\nu_{k,i}\frac{H^\circ_k}{\mathcal{R}T},
\end{eqnarray}
so that
\begin{equation}
    K_{p,i}  = \exp \left( \sum_{k=1}^{N_s}\nu_{k,i}\left(\frac{S^\circ_k}{\mathcal{R}} - \frac{H^\circ_k}{\mathcal{R}T}\right) \right) = \exp \left( \sum_{k=1}^{N_s}\nu_{k,i}\frac{-G^\circ_k}{\mathcal{R}T} \right). \label{eq:K_p}
\end{equation}
where $G^\circ_k$ is the Gibbs free energy obtained by expanding the relations from Eqs.~(\ref{eq:H_RT}) and (\ref{eq:S_R})
\begin{equation}
    \frac{G^\circ_k}{RT} = a_{0,k}(1-\ln T) - \frac{a_{1,k}}{2}T - \frac{a_{2,k}}{6}T^2 - \frac{a_{3,k}}{12}T^3 - \frac{a_{4,k}}{20}T^4 + \frac{a_{5,k}}{T} - a_{6,k}  \label{eq:Ck}
\end{equation}

\subsubsection{Third-body reactions}

Certain reactions require the presence of a third body in order for the reaction to proceed. The third-body concentration is defined with respect to the species concentrations weighted by the third-body efficiencies $\alpha_{k,i}$


\begin{equation}
    [X]_i= \sum_{k=1}^{N_s} \alpha_{k,i} [X_k].
\end{equation}

If all species in the mixture contribute equally as a third body, $\alpha_{k.i}= 1$, and the third-body concentration equals the total mixture concentration 
\begin{equation}
    [X]_i = [M] = \sum_{k=1}^{N_s}[X_k].
\end{equation}
In addition, a single species $m$ can function as a third body, in which case
\begin{equation}
   [X]_i = [X_m].
\end{equation}

\subsubsection{Falloff reactions} \label{sec:fallof}

In contrast to elementary and third-body reactions, the rate constant of falloff reactions depends not only on temperature, but also on pressure. The latter dependence is described as a blending of the constants at low $(k_{0,i})$ and high $(k_{\infty,i})$ pressure, each with corresponding Arrhenius parameters (Eq.~(\ref{eq:arrhenius})). The ratio $k_{0,i}/k_{\infty,i}$ together with the third-body concentration define the reduced pressure $P_{r,i}$
\begin{equation}
    P_{r,i} =
    \begin{cases}
        \begin{aligned}
            &\frac{k_{0,i}}{k_{\infty,i}}[X]_i \quad &&\text{for the mixture as the third body, or}\\
            &\frac{k_{0,i}}{k_{\infty,i}}[X_m] \quad &&\text{for a specific species m as third body.}
        \end{aligned}
    \end{cases}
\end{equation}
The forward rate constant is computed as
\begin{equation}
k_{f,i}(T, P_{r,i})=  k_{\infty,i}\left( \frac{P_{r,i}}{1+P_{r,i}}\right) F_i(T,P_{r,i}),
\end{equation}
where the blending factor $F_i$ is determined on the basis of the Lindemann \cite{Lindemann1922}, Troe \cite{Gilbert1983}, or SRI \cite{Stewart1989} formulations
\begin{equation} \label{eq:blend}
    F_i =
    \begin{cases}
        \begin{aligned}
            &1 \quad &&\text{for Lindemann}\\   
            &F^{(1+(\frac{\log P_r+c}{n-d(\log P_r+c)})^2)^{-1}}_{\text{cent}} \quad &&\text{for Troe, or}\\
            &dT^e\left[ a \, \exp\left(-\frac{b}{T}\right) + \exp\left(-\frac{T}{c}\right) \right]^X \quad &&\text{for SRI.}
        \end{aligned}
    \end{cases}
\end{equation}
For the Troe form, the blending factor is written as 
\begin{equation}
    \log F_i = \left[] 1 + \left ( \frac{\log P_r+c}{n-d(\log P_r+c)} \right)^2 \right]^{-1}\log F_{\text{cent}}, 
\end{equation}
with $c, n$ and $d$ defined as 
\begin{align}
    c &= -0.4 - 0.67\log F_{\text{cent}}, \\
    n &= 0.75 - 1.27\log F_{\text{cent}}, \\
    d &= 0.14, 
\end{align}
and 
\begin{equation}
    F_{\text{cent}} = (1-\alpha)\exp \left( -\frac{T}{T^{***}}\right) +  \alpha \exp \left(-\frac{T}{T^{*}} \right) + \exp \left( -\frac{T^{**}}{T} \right)
\end{equation}
The four parameters $\alpha$, $T^{***}$, $T^{*}$ and $T^{**}$ are specified as inputs. Often, the parameter $T^{**}$ is not used and then the last term of $F_{\text{cent}}$ is omitted. 

In the SRI formulation Eq.~\ref{eq:blend} the exponent $X$ is given by
\begin{equation}
    X = \frac{1}{1+\log^2P_r},
\end{equation}
and $a, b, c$ are supplied parameters, while $d$ and $e$ are optional parameters with default values $d=1, e=0$.

\subsubsection{Pressure dependent reactions} \label{sec:p_log}

For certain reactions, the pressure dependence cannot be adequately described using the modification factor $c_i$ (Eq. (\ref{eq:c_i})) and falloff approach outlined previously. In these cases, an alternative formulation based on logarithmic interpolation between two reference pressures can be employed \citep{Stewart1989, Cantera}. At each reference pressure, the rate constant follows the modified Arrhenius expression:
\begin{eqnarray}
k_1(T) &=& A_1T^{\beta_1} \exp\left(-\frac{E_{\mathcal{R},1}}{T}\right) \quad \text{at} \quad p_1 \quad \text{and} \\
k_2(T) &=& A_2T^{\beta_2} \exp\left(-\frac{E_{\mathcal{R},2}}{T}\right) \quad \text{at} \quad p_2,
\end{eqnarray}
where the Arrhenius parameters $(A_1, \beta_1, E_{\mathcal{R},1})$ and $(A_2, \beta_2, E_{\mathcal{R},2})$ are specified at pressures $p_1$ and $p_2$ respectively.
For any intermediate pressure $p$ between $p_1$ and $p_2$, the forward rate constant can then be determined through logarithmic interpolation:
\begin{equation}
\log k_f(T,p) = \log k_1(T) + (\log k_2(T) - \log k_1(T))\frac{\log p - \log p_1}{\log p_2 - \log p_1}.
\end{equation}

\subsection{Transport properties}


\subsubsection{Pure species viscosity}

Pure species viscosities are determined using the standard kinetic theory expression
\begin{equation}
    \eta_k = \frac{5}{16} \frac{\sqrt{\pi m_k k_b T}}{\pi \sigma^2_{k} \Omega^{(2,2)*}},
\end{equation}
where $\sigma_k$ is the Lennard-Jones collision diameter for the $k-k$  interaction potential, $m_k$ is the mass of molecule $k$, $k_B$ is the Boltzmann constant and $T$ is temperature. The collision integral $\Omega^{(2,2)*}$ depends on the reduced temperature 
\begin{equation}
    T^*_k = \frac{k_b T}{\epsilon_k},
\end{equation}
and the reduced dipole moment  
\begin{equation}
    \delta^*_k = \frac{1}{2}\frac{\mu_k^2}{\epsilon_k \sigma_k^3}
\end{equation}
is expressed in terms of the Lennard-Jones interaction well depth $\epsilon_k$ and the dipole moment $\mu_k$. The value of the collision integral $\Omega^{(2,2)*}$ is determined by a quadratic interpolation of the tables based on the Stockmayer potentials available in \cite{Monchick1961}.

\subsubsection{Binary diffusion coefficients}

Binary diffusion coefficients are functions of pressure and temperature \cite{Kee2017}
\begin{equation}
    D_{jk} = \frac{3}{16}\frac{\sqrt{2 \pi k^3_b T^3 / m_{jk}}}{p \pi \sigma^2_{jk} \Omega^{(1,1)*}}
\end{equation}
where $m_{jk}$ is the reduced molecular mass for the $j-k$ species pair
\begin{equation}
    m_{jk} = \frac{m_jm_k}{m_k+m_k},
\end{equation}
and $\sigma_{jk}$ is the reduced collision diameter. The collision integral $\Omega^{(1,1)*}$ is based on the Stockmayer potentials and depends on the reduced temperature 
\begin{equation}
    T^*_{jk} = \frac{k_BT}{\epsilon_{jk}},
\end{equation}
and the reduced dipole moment
\begin{equation}
    \delta^*_{jk} = \frac{1}{2}\mu^*_{jk},
\end{equation}
where $\epsilon_{jk}$ and $\mu^*_{jk}$ are the reduced interaction well depth and the reduced dipole moment, respectively. Two cases are considered in the computation of the reduced quantities, depending on whether the collision partners are polar or non-polar. If the collision partners are either both polar or both non-polar, the following expressions are used:
\begin{eqnarray}
    \epsilon_{jk} &=& \sqrt{\epsilon_j \epsilon_k}, \\
    \sigma_{jk} &=& \frac{1}{2}(\sigma_j + \sigma_k),\\
    \mu^2_{jk} &=& \mu_j \mu_k.
\end{eqnarray}
When a polar molecule interacts with a non-polar molecule
\begin{eqnarray}
    \epsilon_{np} &=& \xi^2 \sqrt{\epsilon_n \epsilon_p}, \\
    \sigma_{np} &=& \frac{1}{2}(\sigma_j + \sigma_k) \xi^{-\frac{1}{6}},\\
    \mu^2_{np} &=& 0,
\end{eqnarray}
where
\begin{equation}
    \xi = 1 + \frac{1}{4}\alpha^*_n\mu^*_p\sqrt{\frac{\epsilon_p}{\epsilon_n}}.
\end{equation}
In the above expressions $\alpha^*_n$ is the reduced polarizability for the non-polar molecule and $\mu^*_p$ is the reduced dipole moment for the polar molecule: 
\begin{eqnarray}
    \alpha^*_{n} &=& \frac{\alpha_n}{\sigma^3_n}, \\
    \mu^*_p &=& \frac{\mu_p}{\sqrt{\epsilon_p \sigma^3_p}}
\end{eqnarray}

\subsubsection{Pure species thermal conductivity}

The individual species conductivities are composed of translational, rotational and vibrational contributions \cite{Warnatz1982}, 
\begin{equation}
    \lambda_k = \frac{\eta_k}{W_k}(f_{\text{trans}}C_{\upsilon,\text{trans}} + f_{\text{rot}}C_{\upsilon,\text{rot}} + f_{\text{vib}}C_{\upsilon,\text{vib}}),
\end{equation}
where
\begin{eqnarray}
    f_{\text{trans}} &=& \frac{5}{2}\left ( 1 - \frac{2}{\pi} \frac{C_{\upsilon,\text{rot}}}{C_{\upsilon, \text{trans}}} \frac{A}{B}\right), \\
    f_{\text{rot}} &=& \frac{\rho D_{kk}}{\mu_k} \left( 1 + \frac{2}{\pi} \frac{A}{B}\right), \\
    f_{\text{vib}} &=& \frac{\rho D_{kk}}{\eta_k},
\end{eqnarray}
and
\begin{eqnarray}
    A &=& \frac{5}{2} - \frac{\rho D_{kk}}{\eta_k}, \\
    B &=& Z_{\text{rot}} + \frac{2}{\pi}\left( \frac{5}{3} - \frac{C_{\upsilon,\text{rot}}}{R} \frac{\rho D_{kk}}{\eta_k}\right).
\end{eqnarray}
The relationships of the translational, rotational and vibrational contributions to the molar heat capacity at constant volume $C_{\upsilon}$ are different depending on whether the molecule is linear or not. In the case of a linear molecule, 
\begin{eqnarray}
    C_{\upsilon, \text{trans}} &=& \frac{3}{2}R, \\
    C_{\upsilon, \text{rot}} &=& R, \\
    C_{\upsilon, \text{trans}} &=& C_{\upsilon} - \frac{5}{2}R. 
\end{eqnarray}
In the above expressions, $R$ is the universal gas constant. 
For the case of a nonlinear molecule, 
\begin{eqnarray}
    C_{\upsilon, \text{trans}} &=& \frac{3}{2}R, \\
    C_{\upsilon, \text{rot}} &=& \frac{3}{2}R, \\
    C_{\upsilon, \text{trans}} &=& C_{\upsilon} - 3R. 
\end{eqnarray}
In the case of single atoms (e.g. H atoms), there are no internal contributions to $C_{\upsilon}$, and therefore,
\begin{equation}
    \lambda_k = \frac{\eta_k}{W_k}\left( f_{\text{trans}}\frac{3}{2}R, \right),
\end{equation}
where $f_{\text{trans}} = 5/2$. The self-diffusion coefficient is defined as
\begin{equation}
    D_{kk} = \frac{3}{16}\frac{\sqrt{2 \pi k^3_b T^3 / m_{k}}}{p \pi \sigma^2_{k} \Omega^{(1,1)*}}.
\end{equation}
The density is computed from the equation of state for a perfect gas, 
\begin{equation}
    \rho = \frac{pW_k}{RT},
\end{equation}
where $W_k$ is the species molecular weight. 
The rotational relaxation collision number $Z_{\text{rot}}$ is a parameter available at $298 K$ representing the number of collisions required to deactivate a rotationally excited molecule. It is typically a small number of order unity, except for molecules with very small moments of inertia (e.g. $Z_{\text{rot}}$ for H$_2$ is 280). The rotational relaxation collision number has a temperature dependence, for which an expression by Parker~\cite{Parker1959} and Brau and Jonkman~\cite{Brau1970} can be used,
\begin{equation}
    Z_{\text{rot}}(T) = Z_{\text{rot}}(298)\frac{F(298)}{F(T)},
\end{equation}
where, 
\begin{equation} \label{eq:poly_diff}
    F(T) = 1 + \frac{\pi^{3/2}}{2} \left( \frac{\epsilon/k_B}{T} \right)^{1/2} +  \left( \frac{\pi^2}{4} + 2 \right)  \left( \frac{\epsilon/k_b}{T} \right) + \pi^{3/2} \left( \frac{\epsilon/k_B}{T} \right)^{3/2}
\end{equation}


\subsubsection{Polynomial fits of temperature dependence} \label{sec:poly_fit}
To speed up the evaluation of the transport properties, the temperature-dependent parts of the pure species properties are fitted. Instead of evaluating the complex expressions for the properties, only comparatively simple polynomial fits need to be evaluated. In order to avoid costly exponential evaluations later, it is advantageous to use a polynomial fit of the property as a function of the logarithm of temperature. 

For viscosity, 
\begin{equation}
    \eta_k = \sum_{n=1}^{N}a_{n,k}(\ln T)^{n-1}
\end{equation}
and  thermal conductivity,
\begin{equation}
    \lambda_k = \sum_{n=1}^{N}b_{n,k}(\ln T)^{n-1}.
\end{equation}
The binary diffusion coefficients the polynomial fits are computed for each species pair, 
\begin{equation} \label{eq:Dkj}
    D_{kj} = \sum_{n=1}^{N}d_{n,k}(\ln T)^{n-1}
\end{equation}
By default, \texttt{KinetiX} follows the approach of Cantera and uses fourth-order polynomial fits (i.e. $N = 5$), as compared to CHEMKIN which uses third-order polynomials \cite{transport}.

Viscosity and conductivity are independent of pressure, while diffusion coefficients depend inversely on pressure. The diffusion coefficient fits are computed at one standard atmosphere. The subsequent evaluation of a diffusion coefficient is obtained by simply dividing the diffusion coefficients as evaluated from the fit by the actual pressure.

\subsubsection{Mixture-averaged properties}

The mixture-averaged formulation is a compromise between accuracy and computational cost.
The mixture-averaged viscosity is computed using the semi-empirical formula of Wilke~\cite{Wilke1950} modified by Bird, et al.~\cite{Bird2002} 
\begin{equation}
    \eta = \sum_{k=1}^{K}\frac{X_k\eta_k}{\sum_{j=1}^{K}X_j\phi_{kj}},
\end{equation}
where
\begin{equation}
    \phi_{kj} = \frac{1}{\sqrt{8}} \left( 1 + \frac{W_k}{W_j} \right)^{-\frac{1}{2}} \left( 1 + \left( \frac{\eta_k}{\eta_j} \right)^{\frac{1}{2}} \left( \frac{W_k}{W_j} \right)^{\frac{1}{4}} \right)^2.
\end{equation}
and $X_k$ is the molar fraction of species $k$.

The mixture-averaged thermal conductivity can be computed using the combination averaging formula of Mathur et al.~\cite{Mathur1967}, 
\begin{equation}
    \lambda = \frac{1}{2} \left( \sum_{k=1}^{K}X_k\lambda_k + \frac{1}{\sum_{k=1}^{K}X_k/\lambda_k} \right).
\end{equation}

Finally, the mixture-averaged diffusion coefficients for species $k$ is computed as~\cite{Bird2002}, 
\begin{equation} \label{eq:mix_diff}
    D_{km} = \frac{\overline{W} - X_kW_k}{\overline{W}} \left( \sum_{j\neq k}^{K} \frac{X_j}{D_{kj}} \right)^{-1},
\end{equation}
where $\overline{W}=\sum_k X_k W_k$ is the average molecular weight.



\bibliographystyle{elsarticle-num}
\bibliography{refs}


\end{document}